\documentstyle[eqsecnum,aps,psfig]{revtex}

\def\etal{{\it et\thinspace al.}\ }
\def\eion{{(e~+~ion)}\ }
\def\eciv{{(e~+~C~IV)}\ }
\def\civ{{C~IV}\ }
\def\sigdr{{$\sigma_{DR}$}\ }
\def\sigeie{{$\sigma_{EIE}$}\ }

\begin{document}
\draft
\preprint{HEP/123-qed}
\title{Relativistic fine structure and resonance effects in
electron-ion recombination and excitation of (e~+~C~IV)}
\author{Anil K. Pradhan, Guo Xin Chen, Sultana N. Nahar}
\address{
Department of Astronomy, The Ohio State University, Columbus, Ohio
43210\\
}
\author{Hong Lin Zhang}
\address{
Applied Physics Division, Los Alamos
national Laboratory, Los Alamos, NM 87544
}
\date{\today}
\maketitle
\begin{abstract}
 Relativistic close coupling calculations are reported for unified
electronic recombination of \eciv including non-resonant and resonant
recombination processes, radiative and dielectronic recombination (RR
and DR). Detailed comparison of the theoretical unified results with two recent 
experiments on ion storage
rings (Mannervik \etal \cite{metal} and Schippers \etal \cite{setal})
shows very good agreement in the entire measured energy region 2s -- 2p 
with $2pn\ell$ resonances. The results 
benchmark theory and experiments to uncertainties of $\sim$15\%, and
show that the resonant and the background cross sections are
not an incoherent sum of separate RR and DR contributions.
The limiting 
values of the DR cross sections, as $n \rightarrow \infty$, are shown to
correspond to those due to electron impact excitation (EIE) at the 
$^2P^o_{1/2,3/2}$ fine structure thresholds,
delineated for the first time. The near-threshold $2s
^2S_{1/2} - 2p ^2P^o_{1/2,3/2}$ EIE cross sections are also
compared with recent experimental measurements.
The demonstrated threshold fine structure and resonance effects should be
of general importance in excitation and recombination of positive ions.
\end{abstract}


 Although \eion recombination has long been studied experimentally and
theoretically, there appears to be considerable uncertainty over
comparisons between measurements and theory, even for expectedly simple
atomic systems such as C~IV \cite{metal,setal}.
 A comparison of the experimental \eciv DR rates
with theoretical data shows disagreement up to orders of magnitude
\cite{setal}. However, as demonstrated by Mannervik \etal
\cite{metal}, using the ion storage ring CRYRING in Stockholm,
 there are complicated physical effects such as
near-threshold fine structure resonances with unexpectedly large
autoionization widths. 
 Theoretically therefore, it is essential
to account for both the relativistic and the complex electron
correlation and resonance effects accurately.
While experiments measure
the combined cross section for \eion recombination, via the resonances and
the background (since there is no natural separation between the two), 
they are still considered individually as dielectronic and 
radiative recombination (DR and RR) respectively.
Apparently there are difficulties 
in measuring the non-resonant background at very low energies, possibly
owing to external field effects \cite{setal}.
But practical applications generally
require \eion rate coefficients, which in turn require cross sections for
both RR and DR at all relevant energies. To that end a theoretical method
has been developed for an {\it ab initio} unified treatment of both processes, 
based on the close coupling (CC) approximation and its relativistic extension,
the Breit-Pauli R-matrix (BPRM) method 
(e.g. \cite{np92,zp97,pz97,zetal,n00a,n00b}).
The BPRM \eion recombination cross sections for several
ions have been compared with experiments, with excellent agreement in all 
cases \cite{refs}. It is therefore of interest to apply the BPRM method
to elucidate the physical effects and issues related to \eciv recombination,
in direct comparison with experimental data \cite{metal,setal}.
 Dielectronic recombination (DR) is also naturally linked to electron impact
excitation (EIE). At the Rydberg series limit, as $n \longrightarrow \infty $ 
(where the RR
background is negligible), the photon flux in DR should in
principle equal the electron scattering flux at threshold (E = 0), in
accordance with Unitarity \cite{bs,zetal}. Threshold fine structure would
however give rise to related structure in the DR and EIE cross sections.

 In this {\it Letter} we present theoretical calculations based on
the relativistic CC method 
to demonstrate that: (i) the theoretical results for \eion recombination
agree with both ion storage ring measurements \cite{metal,setal} 
to within experimental uncertainties, including near-threshold
resonance strengths and non-resonant background, and (ii) fine structure
resonance series and threshold effects in DR {\it below} the EIE threshold, 
that should
be of general importance but have not heretofore been studied.
The  coupled-channel wavefunction expansion for an \eciv 
may be expressed as

\begin{equation}
\Psi(E; e + \civ) = \sum_{i} \chi_{i}(\civ)\theta_{i}(e) +
\sum_{j} c_{j} \Phi_{j}(C~III),
\end{equation}

where the $\Psi$ denote both the bound (E $<$ 0) and the continuum (E
$>$ 0)
states of C~III, expanded in terms of the core ion eigenfunctions
$\chi_i$(C~IV); the $\Phi_j$ are correlation functions.
The CC approximation, using the efficient R-matrix method
and its relativistic
Breit-Pauli extension \cite{ip,ben95}, enables a solution for
the total $\Psi$, with a suitable expansion over the $\chi_i$.
The extention of the BPRM formulation to unified electronic
recombination \cite{zetal,n00a,n00b} entails the following.
 Resonant and non-resonant electronic recombination takes place into
an infinite number of bound levels of the (e~+~ion) system. These are
divided into two groups: (A) the low-n (n $\leq$ n$_o \approx$ 10)
levels, considered via
detailed CC calculations for photorecombination, with
highly resolved delineation of autoionizing resonances, and (B) the
high-n (n$_o \le n \leq \infty$) recombining levels via DR,
neglecting the background. In previous works (e.g. \cite{zetal})
 it has been shown that
the energy region corresponding to (B), below thresholds for DR,
the non-resonant contribution is negligible.
The DR cross sections converge on to the
electron impact excitation cross section
at threshold (n$\rightarrow \infty$, as
required by unitarity, i.e. conservation of photon and electron fluxes.
This theoretical limit is an important check on the calculations,
 and may also be used to show precisely the behavior of the resonances in 
DR fine structure cross sections as they approach and cross the
fine structure thresholds towards the EIE cross section, as shown
in this work. 

 The BPRM calculations for \eciv recombination involve photorecombination
into 212 low-n levels of C~III, up to $\nu \leq$ 10.0 ($\nu$ is the effective
quantum number), and all $SLJ$ symmetries with $J$ = 0 -- 10 (112 even
parity levels and 110 odd parity levels). In the high-$n$ energy region, $10
< \nu \leq \infty$, the background (RR-type) contribution to \eion
recombination is negligible. We calculate DR cross sections
$\sigma_{DR}$ due to the
resonance series $^2P^o_{1/2} n \ell, ^2P^o_{3/2} n \ell $
approaching the two fine structure thresholds $^2P^o_{1/2,3/2}$, and in
between. Both the
detailed $\sigma_{DR}$ and the resonance averaged $<\sigma_{DR}>$ \cite
{bs,zetal} are computed. 
Finally, the EIE cross sections $\sigma_{EIE}$ are computed
at the $^2P^o_{1/2,3/2}$ thresholds and above.
Details of the calculations will be presented elsewhere,
together with rate coefficients for practical applications.

 Fig.~1(a) shows the detailed unified \eciv recombination cross section
$\sigma_{RC}$ in the $1s^2 2s (^2S_{1/2}) -- 1s^2 2p
(^2P^o_{1/2,3/2})$ region.  In order to compare with experiment, we
compute the rate coefficient $v \cdot \sigma_{RC}$, and convolve 
with a gaussian of $\Delta E$(FWHM)
that corresponds to the experimental resolution in the
heavy-ion storage ring TSR \cite{setal}. Fig.~1(b) shows 
the convolved theoretical results compared with the experimental results 
in 1(c) (Fig.~3 in \cite{setal}). The experimental results in 1(c) (black dots) 
are  reported in the region 2 -- 8.5 eV,
as shown, and compared with theoretical DR results (solid line) in \cite{setal}
(multiplied by a factor of 0.8 and shifted by 0.06 eV).
The present unified $\sigma_{RC}$ in 1(a) show considerably more
detail than the experimental results, but the convolved results agree
remarkably well with the individual $n$-complexes of resonances. We
also incorporate an approximate field ionization cut-off in $<v \cdot
\sigma_{RC}>$, experimentally estimated at $n_F$ = 19, with the results shown 
as the dashed line in 1(b), compared to
the dashed line in 1(c) (the dot-dashed line in 1(c) represents a model
calculation of detection probabilities for high Rydberg states
\cite{setal}). A more accurate ionization cut-off may be possible
by considering overlapping (n,J) manifolds of detailed $\sigma_{RC}$ as
in Fig. 1(a). At the series limit in Fig. 1(b) our results up to 
$n = \infty$ also agree very well with the experimental results
augmented as described in \cite{setal} (shaded portion).

 Although the qualitative and quantitative agreement in Fig. 1 
appears to be excellent,
the present unified results also include the background contribution,
which was measured but subtracted from the reported experimental results. A
very precise quantitative comparison can however be done for the
resonance strength of the $2p 4\ell$ complex measured by both the
CRYRING \cite{metal} and the TSR \cite{setal} experiments.
Fig. 2(a) shows the present detailed unified $\sigma_{RC}$ for the $2p
4\ell$ complex, with the individual resonances identified. As in Fig.1,
the convolved $< v \cdot \sigma_{RC}>$ is shown in Fig. 2(b), and
compared with (i) the CRYRING data (open circles), (ii) TSR data (dark
circles), and (iii) calculated rate by Mannervik \etal \cite{metal} 
(shaded area) that is up to 50\% higher than the experimental values.
We particularly note that our background rate $\alpha_{RC}$ = 
0.2 $\times 10^{-10}$ cm$^3$s$^{-1}$ (dark circle) 
in Fig. 2(b), at E = 0.1 eV, agrees precisely with the measured background
value reported in \cite{setal} at the same energy.
Schippers \etal \cite{setal} quote the measured $2p 4\ell$
resonance strengths of
1.9 $\times 10^{-11}$ eVcm$^3$s$^{-1}$ and 2.5 $\times 10^{-11}$
eVcm$^3$s$^{-1}$ from the CRYRING and the TSR data respectively, a
difference of about 30\%.
Our theoretical value is 2.16 $\times 10^{-11}$ eVcm$^3$s$^{-1}$,  obtained by
direct integration over the resonances in Fig. 2(b), and
subtracting a constant background of 0.2 $\times 10^{-10}$
cm$^3$s$^{-1}$ in the energy region covered by the resonances.
Thus our theoretical value agrees better with each experiment, to $\sim$15\%,
than the two experimental values do with each other, differing by 30\%
(although each experiment has a reported uncertainty of 15\%).

 The present unified results confirm the experimentally
measured background around E $\approx$ 0.1 eV, as reported in Fig. 7 of
\cite{setal}, and in present Fig. 2(b). 
Whereas the experimental data are uncertain at very low
 energies, E $<$ 0.1 eV, due to `excess recombination' possibly due to external
fields, the background may not be so affected at higher energies
E $>$ 0.1 eV. We suggest that, except at energies close to the
RR peak E $\approx 0$, the experiments accurately measure the 
total \eion recombination cross
sections that can, therefore, be directly compared with the unified
theoretical calculations.

 Schipper \etal \cite{setal} do not however report {\it total} \eion
recombination cross section since they  eliminate the measured background.
Instead, they use near-hydrogenic approximations to estimate the
RR-contribution \cite{note} to derive total recombination rates, which
agree with the earlier LS coupling rates of Nahar and Pradhan \cite{np97},
to within experimental uncertainties
at all temperatures except at low-$T\ < 5000K$ (the discrepancy
is due to the omission of K-shell excitation correlation functions $\Phi_j$ (Eq.
1) that leads
to some bound levels of C~III appearing as resonances just at threshold).
However, as seen from Figs. 1 and 2 the \eciv recombination cross
sections may not be considered as an incoherent sum of RR and DR.
The unified calculations on the other hand incorporate the background and
resonant recombination in an {\it ab initio} manner, taking account of
quantum mechanical interference between the RR and DR processes.
We shall compare these approximations in
detail with the present more accurate
BPRM photoionization calculations in the low-energy region 
in a subsequent paper on recombination rates for \eciv.

 Next, we consider the threshold behavior of \eciv DR and EIE.
 In Fig. 3 we delineate the fine structure \sigdr in the energy region
spanned by the fine structure $^2P^o_{1/2,3/2}$ thresholds. Fig. 3(a)
shows the detailed resonances in the vicinity of the two series limits.
Fig. 3(b) shows the \sigdr averaged over the lower resonance series
$^2P^o_{1/2} n \ell$ below the $^2P^o_{1/2}$ level, but still with the
detailed resonance structures due to the higher series $^2P^o_{3/2} n
\ell$ (solid line). The \sigdr averaged over both series is shown as the
dashed line. Above the $^2P^o_{1/2}$, \sigdr is averaged over the $^2P^o_{3/2}
n \ell$ series. The sharp drop in the total \sigdr at the $^2P^o_{1/2}$ 
threshold reflects
the termination of DR due to the $^2P^o_{1/2} n \ell$ resonance series, 
and with the $^2P^o_{3/2} n\ell$ contribution still low in spite of the
fact that $n \approx$ 96. The large drop in the DR cross section
is due to enhanced autoionization
into the excited level, when the $^2P^o_{1/2} n \ell$ channel opens up
at the lower fine structure threshold $^2P^o_{1/2}$ while the radiative 
decay remains constant.
 The $\sigma_{DR}(^2P^o_{3/2} n \ell)$  contribution builds
up to the second peak at $^2P^o_{3/2}$. 

In Fig. 3(b) it is shown that the resonance averaged 
$\lim_{n \rightarrow \infty}
<\sigma_{DR}(^2P^o_{1/2} n \ell)>$  = 242.57 Mb (dark circle at 
$^2P^o_{1/2}$), but the detailed  \sigdr
has resonances due to the higher series $(^2P^o_{3/2} n \ell)$ lying
at and near threshold. The resonance averaged \sigdr at the 
next DR peak, $\lim_{n \rightarrow \infty}
<\sigma_{DR}(^2P^o_{3/2} n \ell)>$  = 441.81 Mb (dark circle at
$^2P^o_{3/2}$). Interestingly, 
the fine structure in the theoretical \sigdr in Fig. 3(a,b) 
appears to be discernible as a small dip in experimental data in 
Fig. 2(c) just below 8 eV.
Although the $^2P^o_{1/2,3/2}$ separation is only 0.013 eV, it may be
possible to detect these fine structure threshold effects in future 
experiments with increased resolution.

 At the 
$^2P^o_{1/2,3/2}$ thresholds the sum of the averaged fine structure
$<\sigma_{DR}>$ = $\sigma_{EIE}$ = 684.38 Mb.
Fig. 3(c) compares the near-threshold
EIE cross sections with the absolute measurements from two recent experiments,
(Greenwood \etal \cite{getal} and Janzen \etal \cite{jetal}), 
convolved over their
respective beam widths of 0.175 eV \cite{getal} and 2.3 eV \cite{jetal}.
Our results are in good agreement with both sets (and
also with another recent experiment by Bannister \etal 
\cite{betal,getal}). Although the present results are the first
CC calculations with relativistic fine structure for \civ, 
their sum is in good agreement with previous LS coupling
CC calculations of $\sigma_{EIE}$ \cite{vmb,grifetal,jetal}.

 In this {\it Letter} we demonstrate several new aspects of \eion recombination and
excitation calculations and experiments: (i) the hitherto most detailed 
unified relativistic CC calculations agree with two sets of experimental data, 
such as to constrain both
theoretical and experimental uncertainties to $\sim$15\%, (ii) 
except close to the RR peak at E $\approx$ 0,
the experiments perhaps need not eliminate the background entirely and 
may report the combined (RR + DR) rate in future, 
(iii) the finely delineated DR resonances
could possibly be used to study field-ionization effects from the 
$n,J$-dependent partial DR cross sections, and (iv)
the fine structure threshold effects in \eciv should
manifest themselves more strongly in heavier and complex ions,
in both DR and EIE.

 This work was partially supported by the National Science Foundation
and the NASA Astrophysical Theory Program. The computational work was
carried out at the Ohio Supercomputer Center.

\def\amp{{Adv. At. Molec. Phys.}\ }
\def\apj{{ Astrophys. J.}\ }
\def\apjs{{Astrophys. J. Suppl.}\ }
\def\apjl{{Astrophys. J. (Lett.)}\ }
\def\aj{{Astron. J.}\ }
\def\aa{{Astron. Astrophys.}\ }
\def\aasup{{Astron. Astrophys. Suppl.}\ }
\def\adndt{{At. Data Nucl. Data Tables}\ }
\def\cpc{{Comput. Phys. Commun.}\ }
\def\jqsrt{{J. Quant. Spectrosc. Radiat. Transf.}\ }
\def\jpb{{J. Phys. B}\ }
\def\pasp{{Pub. Astron. Soc. Pacific}\ }
\def\mn{{Mon. Not. R. Astron. Soc.}\ }
\def\pra{{Phys. Rev. A}\ }
\def\prl{{Phys. Rev. Lett.}\ }
\def\zpds{{Z. Phys. D Suppl.}\ }
\def\adndt{At. Data Nucl. Data Tables}

\newpage
\begin{figure}
\centering
\psfig{figure=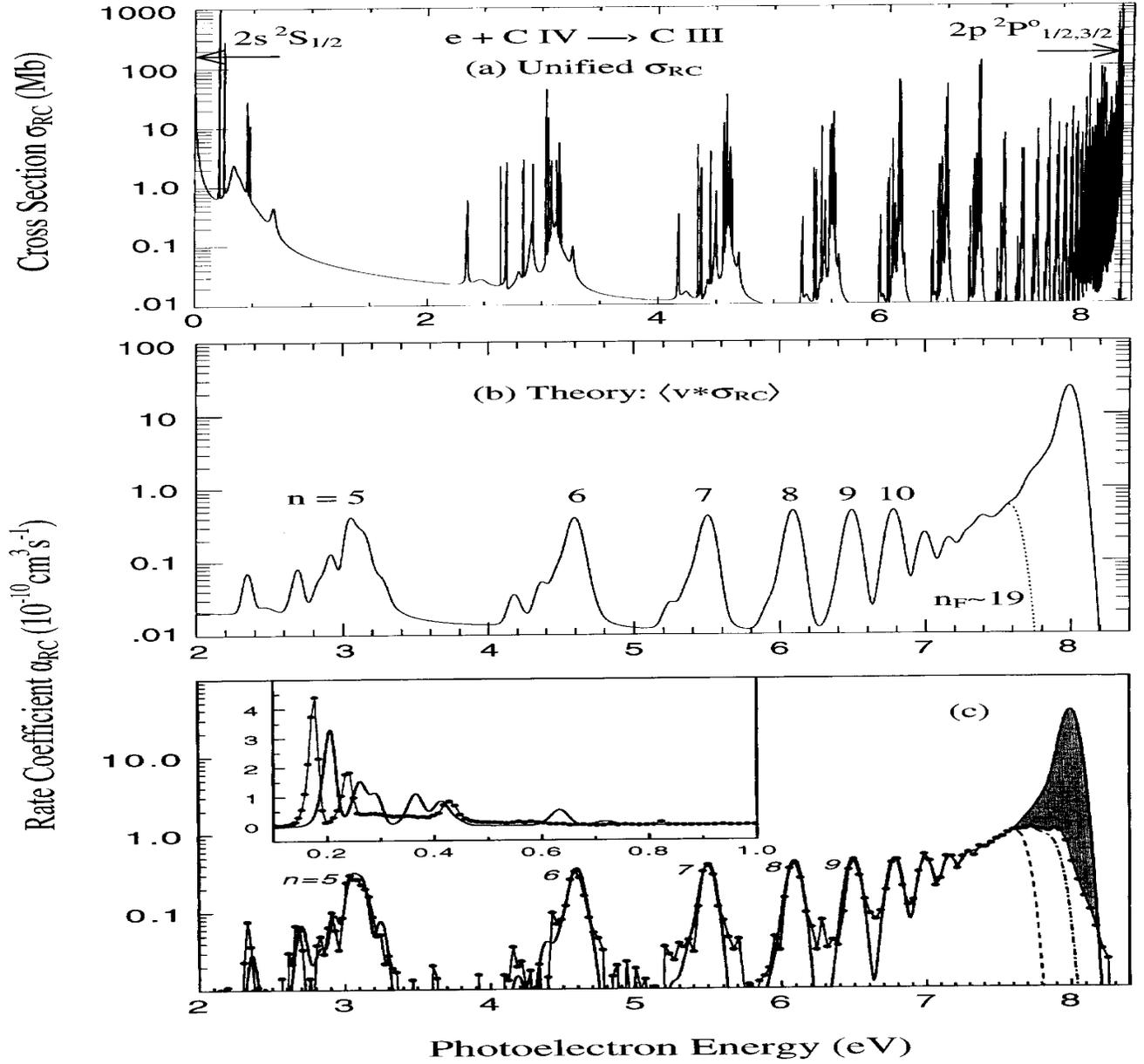,height=17.0cm,width=18.0cm}
\caption{(a) Unified \eciv recombination cross section $\sigma_{RC}$ 
with detailed resonance structures; (b) theoretical rate coefficient 
(v $ \cdot \sigma_{RC}$) convolved
over a gaussian with experimental FWHM \ \protect\cite{setal}; 
(c) the experimentally
measured rate coefficient \ \protect\cite{setal}. The unified $\sigma_{RC}$ in
(a),(b) incorporate the background cross section eliminated from the
experimental data in (c). The dashed and dot-dashed lines represent approximate
field ionization cut-offs (see text).}
\end{figure}

\begin{figure}
\centering
\psfig{figure=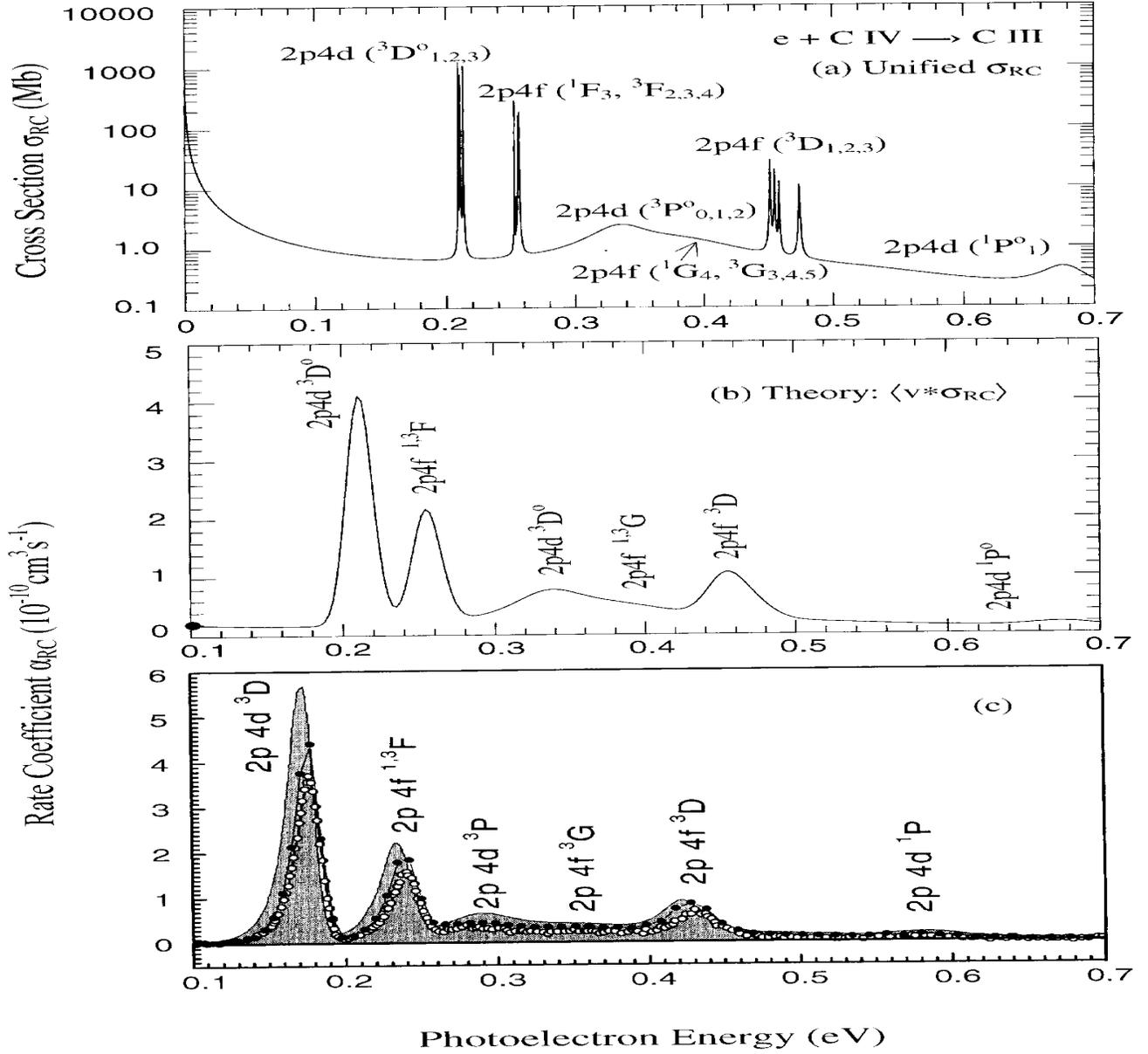,height=17.0cm,width=18.0cm}
\caption{(a) The $2p 4 \ell$ resonance complex: detailed unified
$\sigma_{RC}$; (b) convolved rate coefficient (v $\cdot \sigma_{RC}$); 
(c) experimentally measured values from CRYRING \ \protect\cite{metal} 
(open circles),
TSR \ \protect\cite{setal} (dark circles), and theoretical calculations from 
\ \protect\cite{metal}
(shaded region). The filled circle in (b) at E = 0.1 eV represents the
experimentally measured background values (Fig. 7 in \ \protect\cite{setal}.}
\end{figure}

\begin{figure}
\centering
\psfig{figure=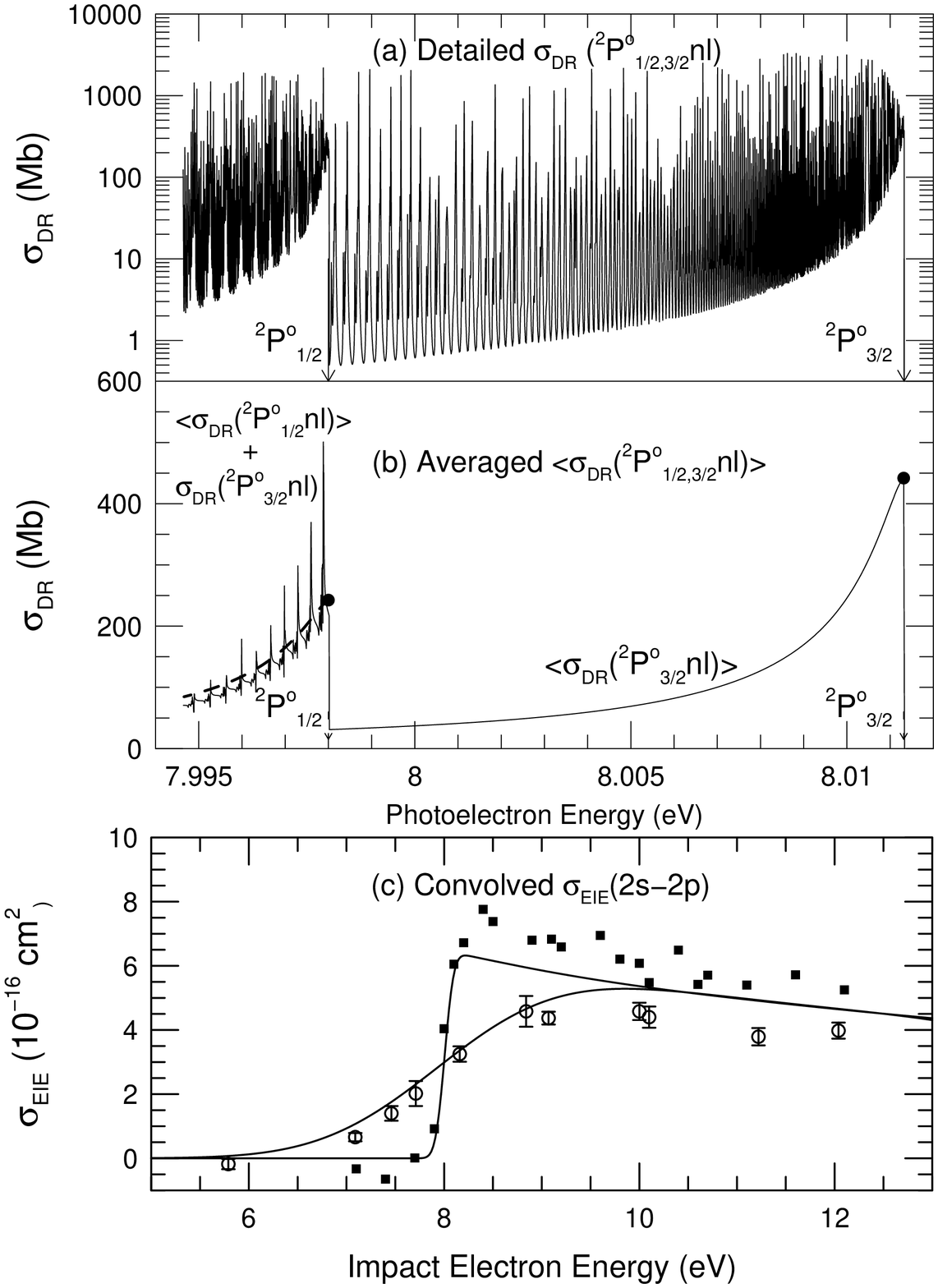,height=17.0cm,width=18.0cm}
\caption{\sigdr and \sigeie of \civ: (a) detailed \sigdr with
$^2P^o_{1/2,3/2} n\ell$ resonances; (b) \sigdr averaged over
$^2P^o_{1/2} n\ell$ and detailed $^2P^o_{3/2} n\ell$ resonances (solid line),
average over the $^2P^o_{3/2} n\ell$ (dashed line); the dark circles
are the peak averaged \sigdr; (c) \sigeie convolved over experimental
data with FWHM = 0.175 eV from \ \protect\cite{getal} (filled squares), and
with FWHM = 2.3 eV from \ \protect\cite{jetal} (open circles).}
\end{figure}


\begin{references}

\bibitem{metal} S. Mannervik, D.R. DeWitt, L. Engstr\"{o}m, J.
Lidberg, R. Schuch and W. Zhong, \prl {\bf 81}, 313 (1998).

\bibitem{setal} S. Schippers, A. M\"{u}ller, G. Gwinner, J. Linkemann,
A. Saghiri and A. Wolf, \apj {\bf 555}, 1027 (2001); the
subtracted background was re-added to the DR contribution to obtain the
total recombination rate coefficient.

\bibitem{np92} S.N. Nahar and A.K. Pradhan, \prl {\bf 68}, 1488 (1992).
\bibitem{zp97} H.L. Zhang and A.K. Pradhan, \prl {\bf 78}, 195 (1997).
\bibitem{pz97} A.K. Pradhan and H.L. Zhang \jpb {\bf 30}, L571 (1997).
\bibitem{zetal} H.L. Zhang, S.N. Nahar, and A.K. Pradhan, \jpb {\bf 32},
1459 (1999).
\bibitem{n00a} S.N. Nahar, A.K. Pradhan, and H.L. Zhang, \apjs {\bf
131}, 375 (2000).
\bibitem{n00b} S.N. Nahar, A.K. Pradhan, and H.L. Zhang, \apjs {\bf 133},
255 (2001).
\bibitem{refs} These include \eion recombination to: C~IV, C~V, O~VII
\cite{zetal}, Ar~XIV \cite{zp97}, Fe~XXIV \cite{pz97}, and Fe~XVII
\cite{petal}
\bibitem{bs} R.H. Bell and M.J. Seaton, \jpb {\bf 18}, 1589 (1985).
\bibitem{petal} A.K. Pradhan, S.N. Nahar, and H.L. Zhang, \apjl {\bf
549}, L265 (2001).


\bibitem{ip} D.G. Hummer, K.A. Berrington, W. Eissner, A.K. Pradhan,
H.E. Saraph and J.A. Tully, Astron. Astrophys. {\bf 279}, 298 (1993).
\bibitem{ben95} K.A. Berrington, W. Eissner, and P. H. Norrington, \cpc
{\bf 92}, 290 (1995).

\bibitem{note} Schippers \etal use RR recombination rates from Pequignot
\etal (D. Pequignot, P. Petitjean, \& C. Boisson, \aa {bf 251}, 680
1991), that are derived from photoionization cross sections by
N. Sakhibulin and A. Willis (\aasup {\bf 31}, 11 (1978)) calculated using
the quantum defect method, which are nearly hydrogenic and differ
from the CC calculations \cite{np97}.

\bibitem{np97} S.N. Nahar and A.K. Pradhan, \apjs {\bf 111}, 339
(1997).
\bibitem{getal} J.B. Greenwood, S.J. Smith, and A. Chutjian, \pra {\bf
59}, 1348 (1999).
\bibitem{jetal} P.H.Janzen, L.D. Gardner, D.B. Reisenfield, D.W. Savin,
and J.L. Kohl, \pra {\bf 59}, 4821 (1999). 
\bibitem{betal} M.E. Bannister, R.-S.Chung, N. Djuric, B. Wallbank, O.
Woiteke, S. Zhou, G.H. Dunn, and A.C.H. Smith, \pra {\bf 57}, 278
(1998).
\bibitem{vmb} V.M. Burke, \jpb {\bf 25}, 4917 (1992).
\bibitem{grifetal} D.C. Griffin, N.R. Badnell and M.S. Pindzola, \jpb
{\bf 33}, 1013 (2000).

\end{references}
\end{document}